\documentclass[aps,prb,showpacs,twocolumn,superscriptaddress,floatfix]{revtex4-2}

\usepackage[colorlinks,bookmarks=false,citecolor=blue,linkcolor=red,urlcolor=blue]{hyperref}
\usepackage{epsfig}
\usepackage{hhline}
\usepackage{graphicx}
\usepackage{enumitem}
\usepackage{dcolumn}
\usepackage{bm}
\usepackage{amsmath,amssymb,mathtools}
\usepackage{placeins}
\usepackage{bbold}


\begin{document}
	\title{Structures of group-15 elemental solids from an effective boundary theory}
	\author{Ashland Knowles}
	\email{gk23dp@brocku.ca}
	\affiliation{Department of Physics, Brock University, St. Catharines, Ontario L2S 3A1, Canada}
	\author{R. Ganesh}
	\email{r.ganesh@brocku.ca}
	\affiliation{Department of Physics, Brock University, St. Catharines, Ontario L2S 3A1, Canada}
	
	\date{\today}

	\begin{abstract}
                We present an effective description for the crystal structures of pnictogen elemental solids. In these materials, each atom contains three valence electrons in $p$ orbitals. They are shared between neighbouring atoms to form valence bonds. We propose a trivalent network model on the simple cubic lattice. As a generalization of a dimer model, we impose a constraint that three dimers must touch every site. We argue that intra-orbital Coulomb repulsion prohibits the formation of two adjacent, parallel dimers. This leads to a tripod-like local configuration at every site. More importantly, it forces every line of the cubic lattice to have alternating dimers and blanks. There is no dynamics as dimers cannot be locally rearranged. A bulk-boundary mapping emerges whereby bonds in the interior are fully described by Ising variables on three bounding planes -- a simple example of holography that may be realized in real materials. 
                To describe the energetics of bonding, we formulate a minimal model in terms of boundary Ising spins. Symmetries reduce the problem to that of three identical, independent, two-dimensional Ising models.   
                An antiferromagnetic Ising-ground-state corresponds to the A7 structure seen in antimony and grey arsenic. An antiferromagnetic phase within a bilayer describes the structure of phosphorene. By stacking such bilayers, we obtain the A17 structure of black phosphorus. The stripe phase of the Ising models describes the cubic gauche structure of nitrogen. As a testable signature, we demonstrate that single impurities will induce long-ranged domain walls.
	\end{abstract}
	
	\maketitle  
	
\section{Introduction}
Crystal structures and the mechanisms that stabilize them are of great interest. In many materials, the observed structure can be fruitfully described as a distortion of an idealized parent structure. A recent example is the family of 1T transition metal dichalcogenides. Starting from an ideal triangular lattice, their structures are obtained by trimerization\cite{Subedi2017,Kojima2023,Chen2018}, a stripe-like 1T$^\prime$ distortion\cite{Brown1966,Zhao2021} or the formation of a ``diamond-chain'' structure\cite{Alcock1965,Lamfers1996,Murray1994,Vitoux2020,SchwarzmüllerStefan2024PSoH}. In this article, we discuss crystal structures of group-15 elemental solids as distortions of a simple cubic lattice. Previous studies have taken the same approach, invoking mechanisms such as Peierls distortion\cite{Iwasaki1986,Burdett1983,Seifert1996,Haussermann2003} and $s$-$p$ orbital mixing\cite{Seo1999}.
This approach can be justified by the fact that two members of this family, arsenic\cite{Beister1990,Tsuppayakorn2018} and phosphorus\cite{Scelta2017,Scelta2018}, exhibit simple cubic phases under pressure.

Group-15 elements or pnictides show various singly-bonded structures in the solid phase. Grey arsenic, antimony, phosphorus (under pressure) and bismuth I exhibit the A7 crystal structure. Black phosphorus, black nitrogen\cite{Laniel2020,Ji2020}, and black arsenic form the A17 crystal structure. 
A third crystal structure of arsenic has been discovered\cite{Matsubara2001} and is characterized as having a structure that spatially alternates between black and grey arsenic structures\cite{Yoshiasa2019}.
Nitrogen forms multiple structures\cite{Yao2021} under pressure, including A7 and cubic gauche\cite{Eremets2004}.
Pnictogens are also known to exhibit various two-dimensional structures\cite{Pumera2017,Hogan2021}. In all these materials, each atom has three valence electrons that reside in $p$ orbitals. They may be shared with neighbouring atoms to form covalent bonds or multicenter bonds\cite{Zahn2025,Golden2017}. With these considerations, we construct a network model in the mould of the Rokhsar-Kivelson dimer model\cite{Rokhsar1988}. Dimers that represent covalent bonds are placed on the links of a cubic lattice. Using simple symmetry-based arguments, we explain the A7, A17 and cubic gauche structures. We build upon our earlier studies of loop\cite{Knowles2025_1} and network models\cite{Knowles2025_2} for transition metal dichalcogenides. In particular, we follow the same paradigm that provides an explanation for the diamond-chain structure\cite{Knowles2025_2} seen in materials such as ReS$_2$, TcS$_2$ and NaMoO$_2$.

An exciting outcome of our approach is a bulk-boundary mapping. The idea of a one-to-one mapping between the bulk state of a system and that of its boundary is well known in the context of quantum gravity. It is described as the holographic principle and is exemplified by the AdS/CFT correspondence\cite{Hubeny2015}. Here, we argue that a simple version of this idea is realized in a family of elemental solids. This suggests that it may be possible to tune bulk properties by perturbing the surface, e.g,. by depositing impurities on the surface of a nano-sized sample. To test the validity of our approach, 
we predict that a single-impurity (e.g., a selenium atom in solid arsenic) will produce long-ranged domain walls -- a proposition that can be readily tested in experiments.  We discuss exciting consequences such as restricted defect mobility and fracton-like character.

\section{Bonding and Configurations of Dimers}
\label{sec.configspace}
\begin{figure}
\includegraphics[width=0.7\columnwidth]{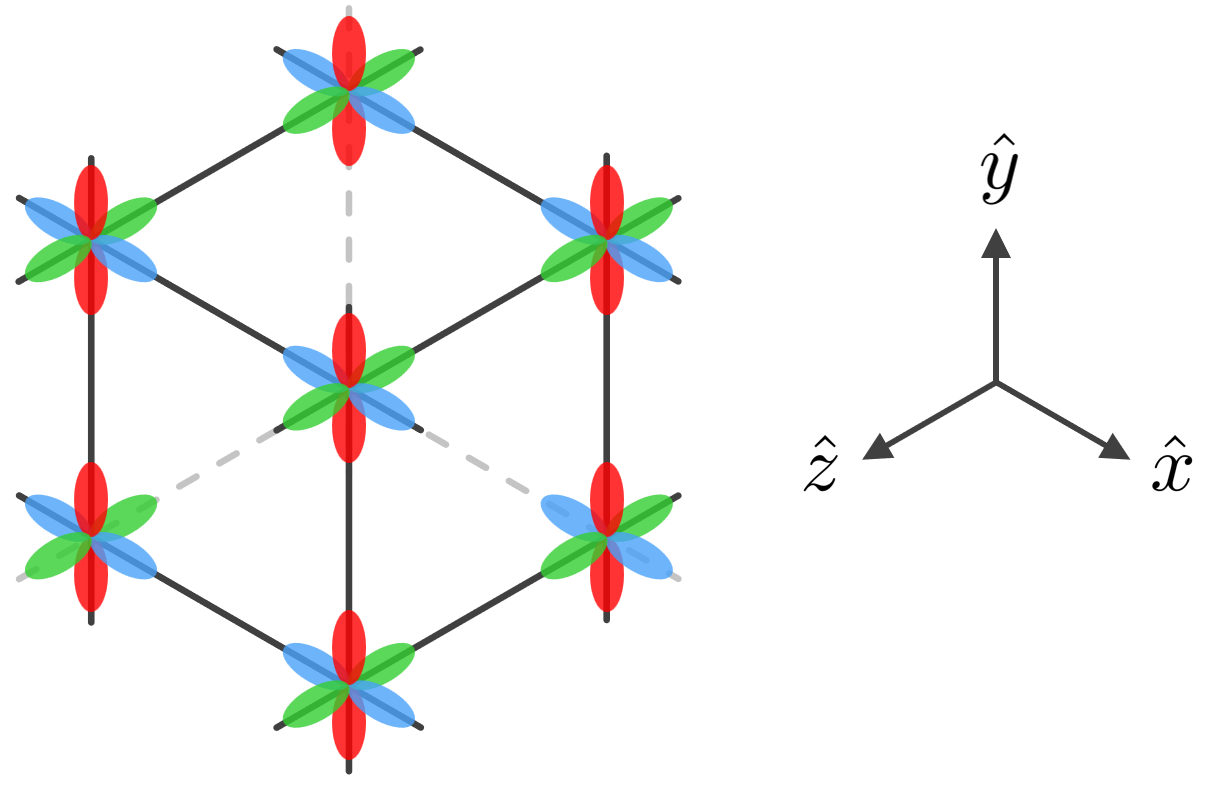}
\caption{A simple cubic lattice viewed along the body diagonal of a cube. Three $p$ orbitals are shown at each site, with $p_x$ orbitals oriented along the $x$-axis, $p_y$ orbitals along the $y$-axis, and $p_z$ orbitals along the $z$-axis.}
\label{fig.Orbitals}
\end{figure}

\begin{figure}
\includegraphics[width=0.8\columnwidth]{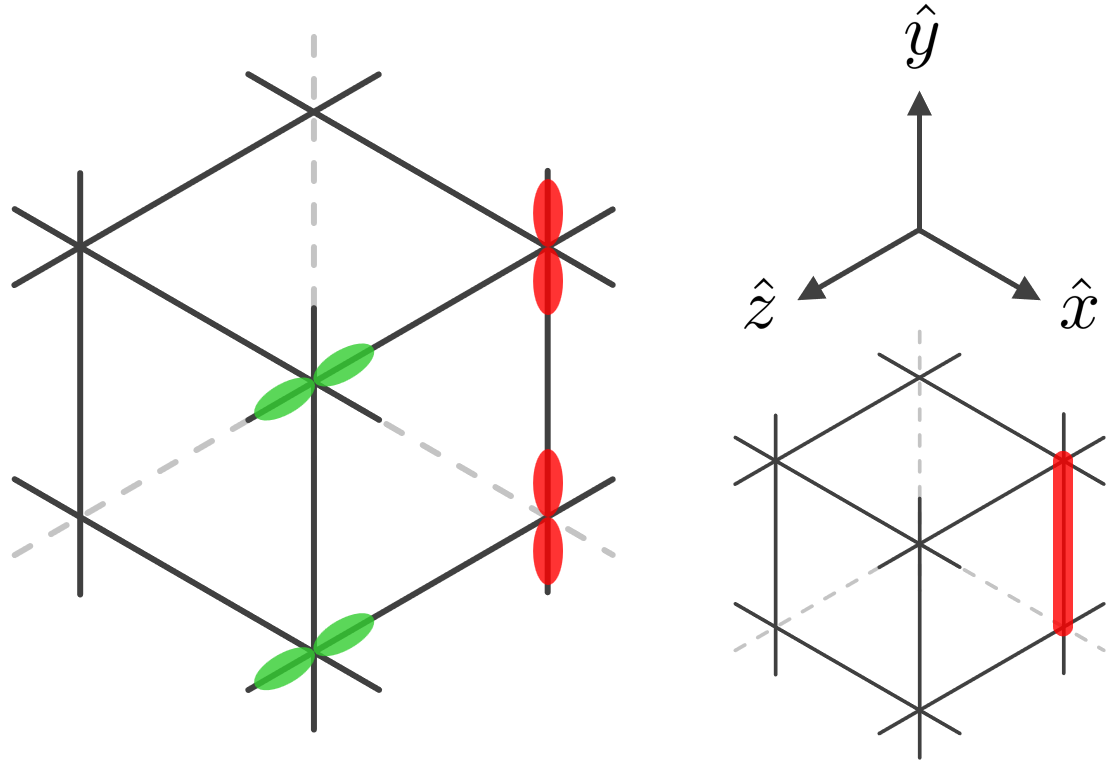}
\caption{Overlaps between $p$ orbitals. The $p_z$ orbitals in green are on neighboring sites, but their overlap is weak since the line connecting them is not along the $z$-axis. They can, at best, form a weak $\pi$ bond. A $p_z$ orbital and its neighbouring $p_y$ orbital have zero overlap due to symmetry. 
The $p_y$ orbitals in red are also on neighboring sites. They have a strong overlap, since the line connecting them is along the $y$-axis. They can form a strong $\sigma$ bond. When there is an electron in each of the two overlapping $p_y$ orbitals, they form a covalent bond which we represent with a dimer, as shown on the right.}
\label{fig.Overlap}
\end{figure}

We begin with an idealized structure where atoms of a group-15 element reside at the sites of a simple cubic lattice. The valence $p$ orbitals lie along three perpendicular axes $x$, $y$ and $z$ that point towards neighbouring sites, see Fig.~\ref{fig.Orbitals}. 
This geometry leads to strong overlaps between $p_x$ orbitals along $x$-bonds, $p_y$ orbitals along $y$-bonds and $p_z$ orbitals along $z$-bonds -- as showin in Fig.~\ref{fig.Overlap}. This allows for orbital-specific covalent bonds between nearest neighbours. Reference~\cite{Knowles2025_2} discusses a similar scenario on the  two-dimensional triangular lattice, based on the geometry of $t_{2g}$ orbitals. Here, we extend this approach to the three-dimensional cubic lattice, based on $p$-orbital geometry.

To describe bonding in group-15 solids, we place dimers on the nearest-neighbour bonds of the simple cubic lattice based on two rules:
\begin{itemize}[leftmargin=*]
    \item Trivalency rule: Exactly three dimers must touch every site of the lattice.
    \item Bending constraint: Two dimers attached to the same site cannot be parallel.
\end{itemize}
The dimers represent covalent bonds. The trivalency rule reflects the fact that each atom hosts three valence electrons that are shared with neighbors. With three dimers at each site, we obtain a network of dimers on the cubic lattice. We call this a trivalent network as the coordination number (number of connected neighbours at each node) is three.

The bending constraint arises from intra-orbital Coulomb repulsion. If a site has bonds along the same direction (e.g., a bond along $+x$ and another along $-x$), both bonds will have the same orbital character. At this site, two electrons must reside in the same orbital ($p_x$ in this example) to form two parallel bonds. The resulting steep Coulomb energy cost will effectively forbid such configurations. This can be viewed as an application of Hund's rule to atoms within the solid.
We imagine a sequence of contiguous bonds as a line passing through sites. As successive bonds cannot be parallel, the line must ``bend'' at each site.

There are many ways to place dimers on the cubic lattice subject to the two rules. We designate each of these possibilities as a valid network configuration. The following two corollaries result from the two previous rules:

   \begin{figure}
    \includegraphics[width=\columnwidth]{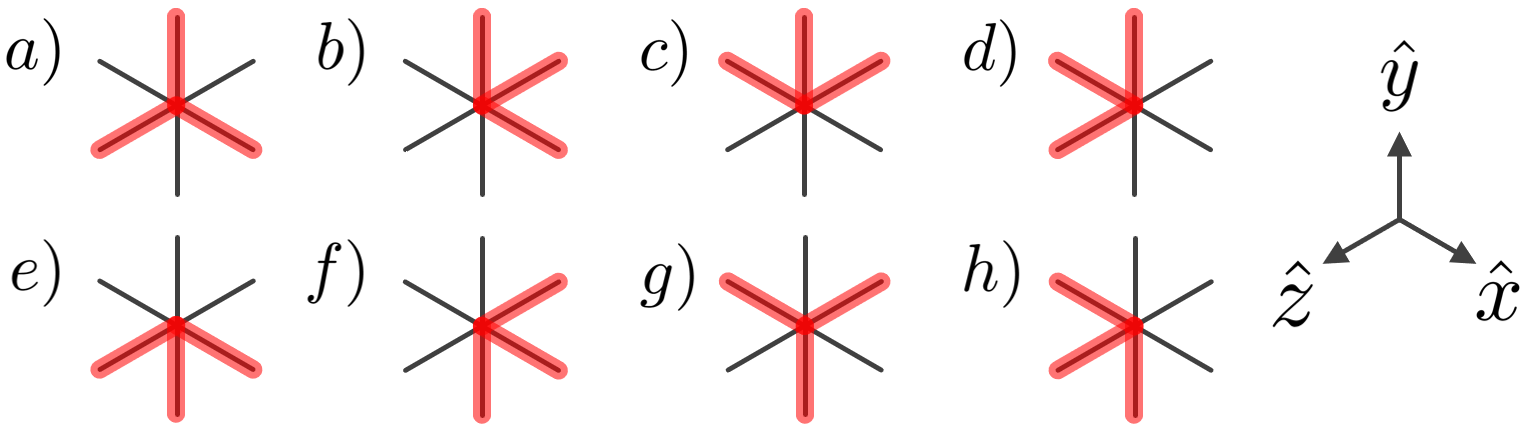}
    \caption{Eight possible orientations of a tripod at each site. The tripod consists of three dimers attached to the site at the centre.}
    \label{fig.Tripods}
    \end{figure}

\begin{figure*}
    \includegraphics[width=2\columnwidth]{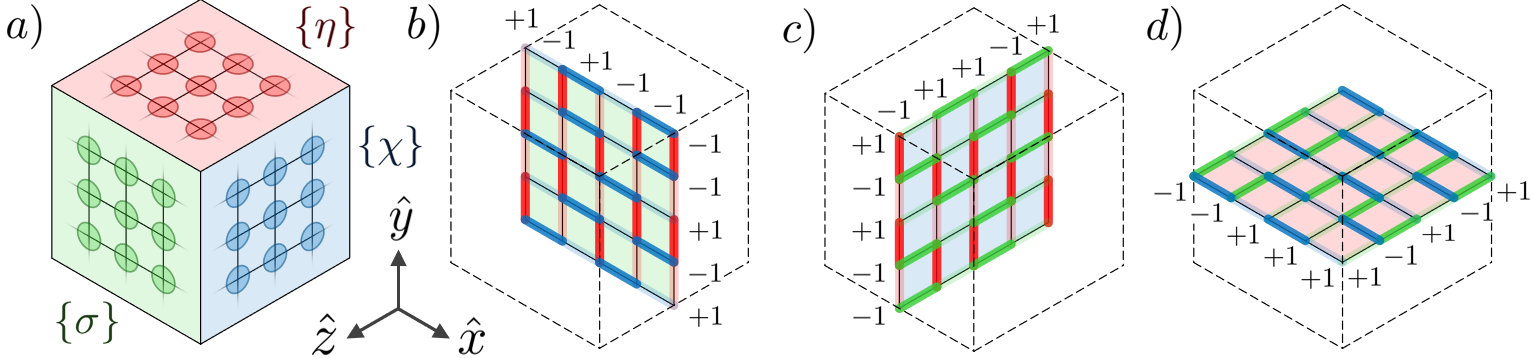}
    \caption{a) A simple cubic lattice with open boundaries. 
    Boundary variables are defined on square grids on three bounding surfaces. Each boundary variable determines dimer placements along a line that is directed into the cube. The $xy$ plane is colored green and has variables $\{\sigma\}$, the $yz$ plane is colored blue and has variables $\{\chi\}$, and the $zx$ plane is colored red with variables $\{\eta\}$.
    Boundary variables are Ising-like, taking values $+1$ or $-1$. b) A plane of dimers with $\chi$ and $\eta$ variables at the boundaries. Dimers fixed by the $\chi$ variables (shown in blue) are perpendicular to the $yz$ plane. Similarly, dimers fixed by $\eta$'s (shown in red) are perpendicular to the $zx$ plane. The same convention is used to highlight dimers in (c) and (d). c) A plane of dimers with $\sigma$ and $\eta$ variables at the boundaries. d) A plane of dimers with $\sigma$ and $\chi$ variables at the boundaries.}
    \label{fig.IsingCube}
    \end{figure*}

\begin{enumerate}[leftmargin=*, itemsep=0pt]
    \item The three dimers that touch at a site must be mutually perpendicular. This reflects the fact that $p$ orbitals lie along the $x$-, $y$- and $z$-axes (see Fig.~\ref{fig.Orbitals}). We cannot have two parallel dimers at a site. We also cannot have two parallel blanks, as three dimers cannot fit on the remaining bonds without violating the bending constraint. With three mutually perpendicular dimers at each site, we form a `tripod'. Each site may have one of eight orientations for its tripod (see Fig.~\ref{fig.Tripods}).  
    \item Every straight line of the cubic lattice must have an alternating sequence of dimers and blanks. For example, if a bond along $x$ hosts a dimer, the adjacent $x$-bond must be blank. Conversely, if an $x$-bond is blank, the next must host a dimer. This can be seen from the geometry of tripods in Fig.~\ref{fig.Tripods}.
    It follows that each line must be in one of two configurations. We may fix one particular bond to be either a dimer or a blank; this immediately fixes all other bonds on the line.
\end{enumerate}
These corollaries strongly constrain dimer-dimer couplings. They also provide a way to enumerate network configurations and to formulate a boundary theory.

\section{Boundary Theory}
\label{sec.BoundaryTheory}
    \subsection{Network configurations}

    We consider a simple cubic lattice with open boundaries as shown in Fig.~\ref{fig.IsingCube}. We consider three bounding surfaces: one parallel to the $xy$ plane, one parallel to the $yz$ plane and a third parallel to the $zx$ plane. On each surface, we define a square grid of points where the surface intersects lines of the cubic lattice. In light of the second corollary in Sec.~\ref{sec.configspace}, we define three families of Ising variables, $\{\sigma\}$, $\{\chi\}$ and $\{\eta\}$ -- one on each square grid.
    The two values of each Ising variable ($\pm 1$) represent the two possible dimer configurations on the normal line emanating from the point. That is, each Ising variable determines the placement of dimers on a line directed into the cube (with alternating dimers and blanks). Figs.~\ref{fig.IsingCube}(b)-(d) shows examples of boundary Ising variables and the resulting dimer arrangements in the interior.

    Since every bond in the bulk exists on a line of alternating dimers and blanks, the bulk configuration of dimers is completely determined by boundary variables. This is a holographic construction similar to the one discussed in Ref.~\onlinecite{Knowles2025_2}. This construction provides a way to enumerate all valid bonding configurations in the bulk. With an $L_x \times L_y \times L_z$ cluster, the number of Ising variables on the $xy$ plane is $(L_x \times L_y)$, that on the $yz$ plane is $(L_y \times L_z)$ and that on the $zx$ plane is $(L_z \times L_x)$.
    The number of distinct dimer arrangements is the number of Ising configurations, given by
    \begin{equation}
        N(L_x,L_y,L_z)=2^{(L_xL_y+L_yL_z+L_zL_x)}.
        \label{eq.N_c}
    \end{equation}
    As seen from this expression, the configuration space grows exponentially with system size.
    
    \subsection{Energy in the model}
    We next build a minimal Hamiltonian in the spirit of the Rokhsar-Kivelson model\cite{Rokhsar1988}. Below, we present arguments for large system sizes, neglecting small corrections that may arise from peripheries. We may associate kinetic energy with local rearrangements of dimers. However, the structure of the configuration space here does not allow local rearrangements. From the second corollary in Sec.~\ref{sec.configspace}, a valid configuration must have an alternating sequence of dimers and blanks along every line of the lattice. Any local rearrangement will necessarily violate this requirement. The simplest allowed rearrangement corresponds to shifting dimers along an entire line of the cubic lattice. Such long-ranged dynamical processes can be neglected, as a large number of dimers must be repositioned. This leaves us with a dynamics-free description, analogous to a classical dimer model.

\begin{figure}
    \includegraphics[width=1\columnwidth]{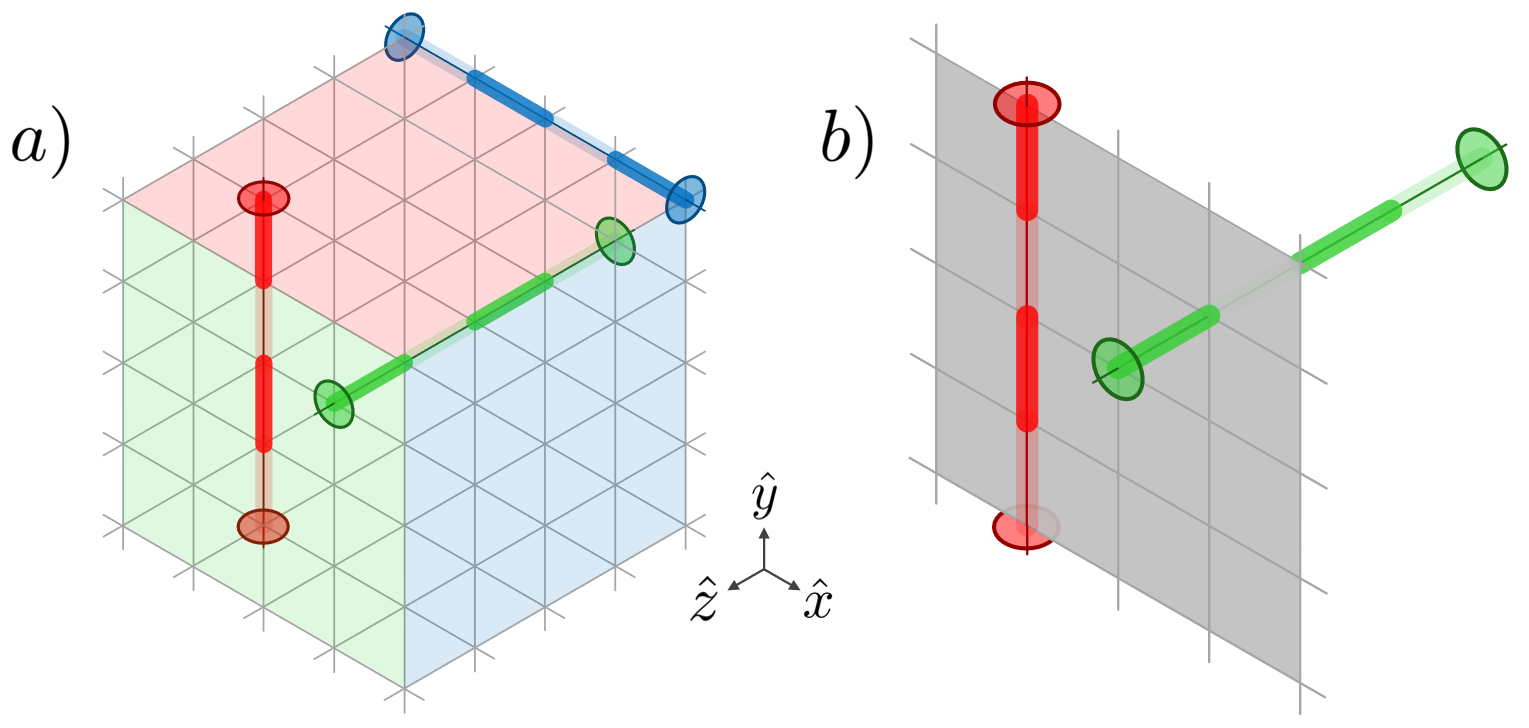}
    \caption{a) Perpendicular lines on the cubic lattice. Each line contains an alternating sequence of dimers and blanks. Dimers are shown in red, blue and green on a $5 \times 5\times 5$ section of a large cubic lattice. Shaded circles mark the ends where the line intersects the boundary of the $5\times 5\times 5$ section. The sides of the cube that are nearest to the viewer are transparent and are shaded in red, green, and blue. b) A plane from (a) that contains the line of red dimers is highlighted. The line of green dimers from (a) intersects this plane and is perpendicular to it. A mirror reflection about this plane leaves the red dimers unchanged, but shifts the green dimers. 
    }
    \label{fig.PE_perp}
    \end{figure}

\begin{figure}
    \includegraphics[width=0.8\columnwidth]{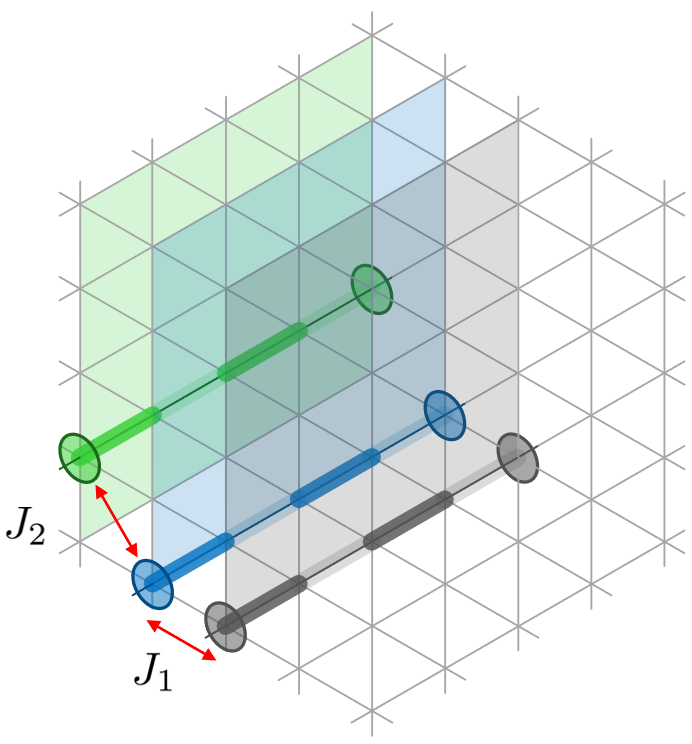}
    \caption{Parallel lines of alternating dimers and blanks. The line of grey dimers and blanks is a nearest-neighbor of the line of blue dimers and blanks. The coupling between these two lines is taken to be $J_1$. The line of green dimers and blanks is a second-nearest-neighbor of the line of blue dimers and blanks. The coupling between these lines is taken to be $J_2$.}
    \label{fig.J1J2}
    \end{figure}

    We next consider potential energy arising from pairwise dimer-dimer interactions. Each dimer is part of a line that contains an alternating sequence of dimers and blanks. As a result, we consider interactions between pairs of lines at once, rather than pairs of individual dimers. If we choose two lines of the cubic lattice, they may either be (i) perpendicular to one another, or (ii) parallel. We consider these cases separately below. Examples of perpendicular lines of dimers are shown in Fig.~\ref{fig.PE_perp}, while examples of parallel lines of dimers are shown in Fig.~\ref{fig.J1J2}.

    As an example of a pair of perpendicular lines, we may consider one line along $y$ and another along $z$ as shown in Fig.~\ref{fig.PE_perp}(b). As each line has two possible dimer arrangements, we have four configurations for the pair of lines. Crucially, these four configurations are related to one another by symmetry. For example, we may take a mirror reflection about an $xy$ plane of the cubic lattice. This is a symmetry operation that does not change the energy of the system. Under this operation, dimers on the $y$-line are unchanged. However, dimers and blanks on the $z$-line are interchanged. This demonstrates that changing the configuration on the $z$-line does not change the potential energy. Similarly, a mirror reflection about an $xz$-plane ensures that a shift in the $y$-line does not change the energy.  We conclude that pairwise interactions between any pair of perpendicular lines yield a constant energy contribution that does not vary with dimer positions. Such contributions can be neglected as they merely provide an overall shift to the energy. In the boundary description, a pair of perpendicular lines is represented by two Ising variables that belong to distinct families. For example, lines along $y$ and $z$ are represented by Ising variables from $\{\eta\}$ and $\{\sigma\}$ respectively. 
    We conclude that the three Ising families, $\{ \sigma\}$, $\{ \chi\}$ and $\{\eta\}$, are decoupled with no interaction across families.


    We next consider parallel lines in the lattice. Boundary Ising variables are assigned as described in Fig.~\ref{fig.IsingCube}. With two possible configurations for each line, we have four configurations for a pair of lines. Each line is associated with an Ising variable defined at its extremity. If both lines carry the same Ising value, their dimer configurations are related by a simple translation in the transverse direction, see Fig.~\ref{fig.J1J2}. The case where both Ising variables are $+1$ has the same energy as when both are $-1$, as these situations are related by a unit translation in the direction of the lines. If the Ising variables are opposite, dimers on the two lines are staggered with respect to one another. Due to translational symmetry, we have the same energy regardless of which Ising variable is chosen to be $+1$ (with the other being $-1$). 
    Based on these arguments, the  
    potential energy takes the form of a simple Ising coupling between boundary variables of the same family, e.g., $J_{ij} \sigma_i \sigma_j$.

    By translational symmetry, the Ising coupling can only depend on the distance between spins (on the boundary). For example, each $\sigma$ variable has four nearest neighbours within the plane. The coupling to each of the four neighbours must be the same. On physical grounds, the coupling must decay with distance. As a minimal model, we restrict couplings to nearest and next-nearest neighbours. Examples of nearest-neighbor and second-nearest-neighbor spins are shown in Fig.~\ref{fig.J1J2}.

    Based on these arguments, we arrive at an expression for energy in terms of the boundary Ising variables, 
    \begin{eqnarray}
        E&=&J_1 \sum_{\langle i,j \rangle}\sigma_i\sigma_j + J_2\sum_{\langle\langle k,l \rangle\rangle}\sigma_k\sigma_l\nonumber  \\
        &+& J_1 \sum_{\langle i,j \rangle}\chi_i\chi_j \nonumber  + J_2\sum_{\langle\langle k,l \rangle\rangle}\chi_k\chi_l \\
        &+& J_1 \sum_{\langle i,j \rangle}\eta_i\eta_j + J_2\sum_{\langle\langle k,l\rangle\rangle}\eta_k\eta_l.
        \label{eq.E}
    \end{eqnarray}
    Due to cubic rotational symmetry, the same coupling constants, $J_1$ and $J_2$, operate on each surface. Eq.~\eqref{eq.E} describes three independent, but identical, spin systems, each being a $J_1$-$J_2$ square-lattice Ising model. Their spin states determine the bulk network configuration.
    We note that the coupling constants $J_1$ and $J_2$ must scale with linear system size. On an $L \times L \times L$ cubic lattice, the energy of the bulk must scale as $L^3$ to be an extensive quantity. The expression in Eq.~\eqref{eq.E}, however, describes three square-lattices with energy scaling as $J_{1/2} L^2$. In order to reconcile these scaling relations, $J_1$ and $J_2$ must scale linearly with $L$.
    In the next section, we use Eq.~\eqref{eq.E} to obtain phases of network configurations.

\section{Phases}
\label{sec.phasediagram}
    We use the well-known phase diagram for the ground state of the classical $J_1$-$J_2$ Ising model on the square lattice\cite{Fan1969,Barber1979,Oitmaa1981}. There are three ground-state phases: ferromagnetic, antiferromagnetic and stripe.
    The ferromagnetic and antiferromagnetic phases have a two-fold degeneracy, arising from a global spin-flip operation (interchanging $+1$ and $-1$). The stripe phase has a degeneracy of four. Apart from a global spin flip, we have an additional two-fold choice between horizontal and vertical stripe directions [see Figs.~\ref{fig.Stripes}(a,b)]. Since the same $J_1$ and $J_2$ apply to each surface, all three surfaces will be in the same qualitative phase. However, on each surface, we may independently pick one of the degenerate ground states. 
    In turn, this can give rise to multiple network configurations in the bulk. Below, as a function of $J_1$ and $J_2$, we enumerate all symmetry-distinct ground states of the network model. If two network configurations can be obtained from one another by translation, reflection or 
    rotation, we only keep one.

    In order to compare network configurations with observed crystal structures, we introduce distortions in the cubic lattice as follows. Starting from the ground-state network configuration, we identify the tripod orientation at each site -- see Fig.~\ref{fig.Tripods}. We shift each site by a fixed distance towards the base of its tripod. For example, if a certain site is attached to dimers along the $+x$, $+y$ and $+z$ directions, we shift the position of this site towards the $\hat{x}+\hat{y}+\hat{z}$ direction.  
    This amounts to decreasing the bond-length on bonds that contain a dimer and increasing the bond-length on blanks. Below, we present figures with distortions introduced in this manner.

    \begin{figure}
        \includegraphics[width=1\columnwidth]{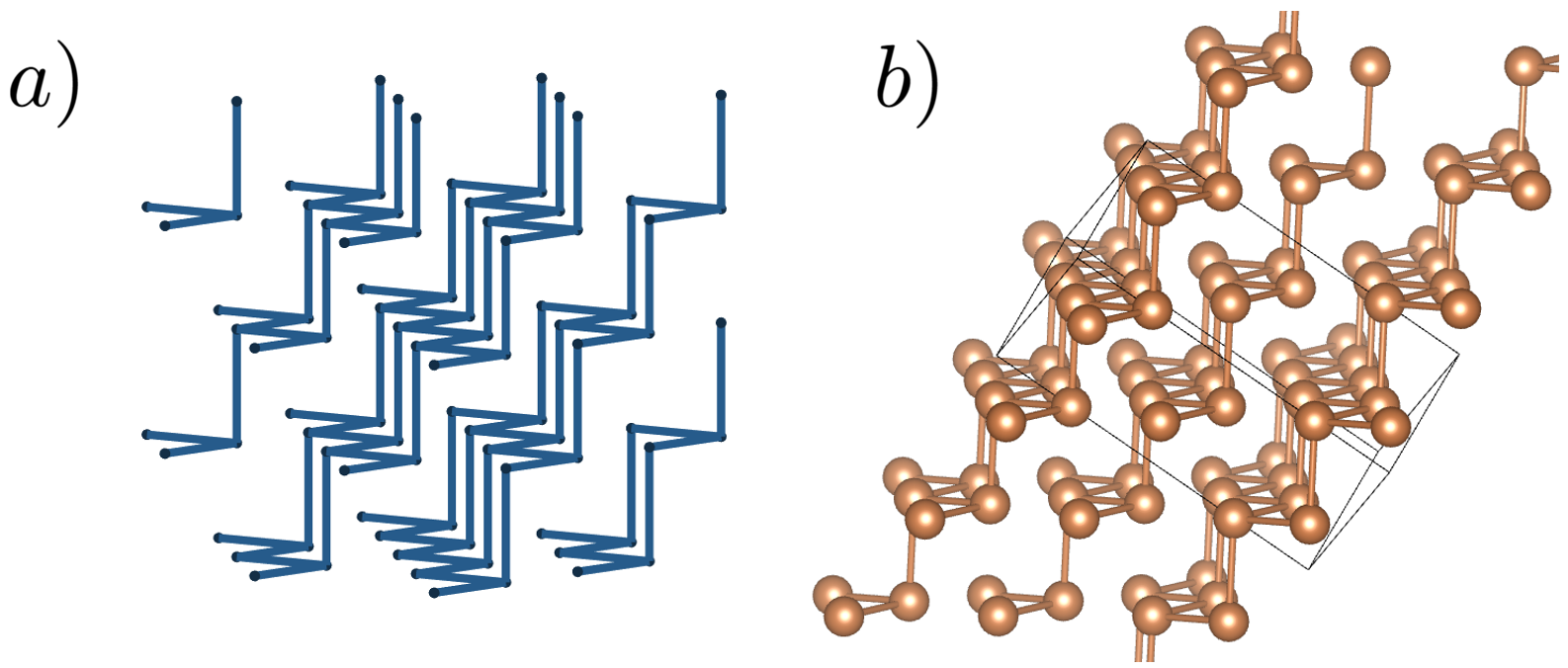}
        \caption{$a)$ Network configuration formed when each surface-Ising model  is in the antiferromagnetic phase. Dimers are shown in blue. They correspond to shorter bond lengths in the elemental solid. $b)$ The A7 structure in antimony\cite{Schiferl1981}.}
        \label{fig.J1AntiF}
    \end{figure}

    \begin{enumerate}[label=(\roman*),wide, labelwidth=!, labelindent=0pt]
        \item \textbf{Antiferromagnetic phase:} The Ising model yields an antiferromagnetic phase when $J_1$ is positive and dominant. There are two possibilities for each surface, related by swapping $+1$'s and $-1$'s. With three boundary surfaces, this naively leads to a degeneracy of $2^3=8$. However, these eight network configurations are symmetry-related, giving rise to a unique network-ground-state. 
        This state corresponds to the A7 crystal structure as shown in Fig.~\ref{fig.J1AntiF}, realized in antimony. 
        This structure naturally forms separable layers. When a single layer is exfoliated, it has the structure of single-layer  antimony, or antimonene\cite{Ares2016,Assebban2020}, and blue phosphorene\cite{Zhu2014,Xu2017}.

        \item \textbf{Ferromagnetic phase:} When $J_1$ is negative and dominant, all Ising variables on a surface take the same value. 
        We have two ground states on each surface, corresponding to all Ising variables being $+1$ or all being $-1$. Naively, with three surfaces, this leads to a degeneracy of $2^3=8$. However, all eight network configurations are related by translations. We have a unique minimum-energy network that consists of `breathing' cubes as shown in Fig.~\ref{fig.BreathingCubes}. As far as we are aware, this structure has not been observed in any pnictogen material. However, in the limit of strong covalent bonding, tightly-bound cubes may separate from the solid to form independent molecules. Such a cubic molecular structure has been proposed for nitrogen, with the name octaazacubane\cite{Lauderdale1992,Glukhovtsev1996}. This hypothetical molecule is believed to store a large amount of potential energy in its bonds, with potential use as an explosive. 
        \begin{figure}        
        \includegraphics[width=0.6\columnwidth]{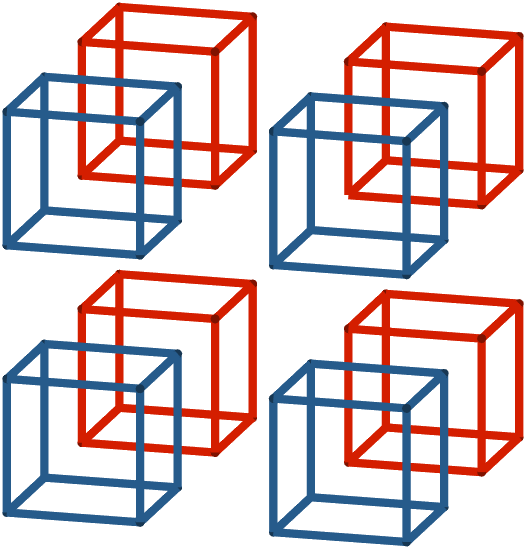}
            \caption{The breathing-cube structure formed by dimers in the ferromagnetic phase. Cubes nearest to the viewer are shown in blue, while the farther ones are in red.}
            \label{fig.BreathingCubes}
        \end{figure}
        \begin{figure}
            \includegraphics[width=1\columnwidth]{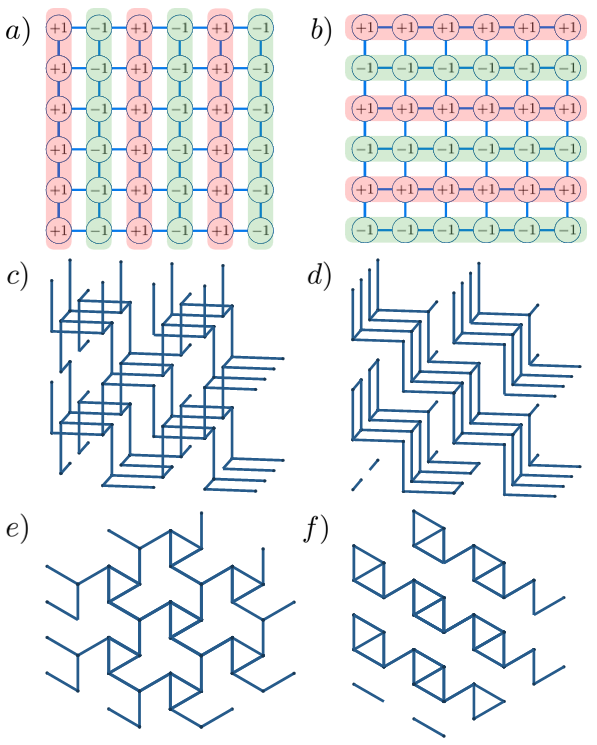}
            \caption{Two orientations of stripes are shown in (a) and (b). Each orientation allows for two configurations that are related by a global spin flip.
            Two symmetry-distinct network configurations arise, as shown in (c) and (d). Viewing the configuration in (c) along a body diagonal of the cubic lattice gives the pattern shown in (e). Viewing the configuration in (d) along a body diagonal of the cubic lattice gives the pattern shown in (f).}
            \label{fig.Stripes}
        \end{figure}   
        
        \item \textbf{Stripes:} This phase appears when $J_2 > \vert J_1\vert/2$. On each surface, Ising variables form stripes as shown in Figs.~\ref{fig.Stripes}(a,b). There are four possible ways for a surface of Ising variables to form stripes. With three surfaces, the naive degeneracy of network configurations is $4^3=64$. Upon examination, we find only two unique configurations. They are shown in Figs.~\ref{fig.Stripes}(c-f) (see Appendix~\ref{app.lines} for an illustration with details). The two configurations in Fig.~\ref{fig.Stripes}(c) and (d) cannot be related by symmetry operations. For instance, there are `squares' (closed loops with four dimers) in Fig.~\ref{fig.Stripes}(d), but not in Fig.~\ref{fig.Stripes}(c). 

        Remarkably, the configuration in Fig.~\ref{fig.Stripes}(c) corresponds to the cubic gauche structure of nitrogen\cite{Eremets2004,Benchafia2017,Laniel2019}. This can be seen from Fig.~\ref{fig.cg_N} which shows the structure of Fig.~\ref{fig.Stripes}(c) with a concomitant distortion. The resulting structure shows excellent agreement with the observed cubic gauche structure of nitrogen. Phosphorus and arsenic have also been predicted to be stable in the cubic gauche (also called $K_4$) structure\cite{Liu2016,Wang2021}. 

        The configuration in Fig.~\ref{fig.Stripes}(d) is layered. We call this the square-octagon structure as it contains four-membered and eight-membered rings. 
        If a single layer is separated by exfoliation, this yields the square-octagon or haeckelite structure that has been proposed for phosphorene\cite{Guan2014,Barik2021,Regragui2025} and other pnictogens\cite{Ersan2016,Carrete2017}. This structure is shown Fig.~\ref{fig.Stripes_SO} with an appropriate distortion.

        \begin{figure}
            \includegraphics[width=1\columnwidth]{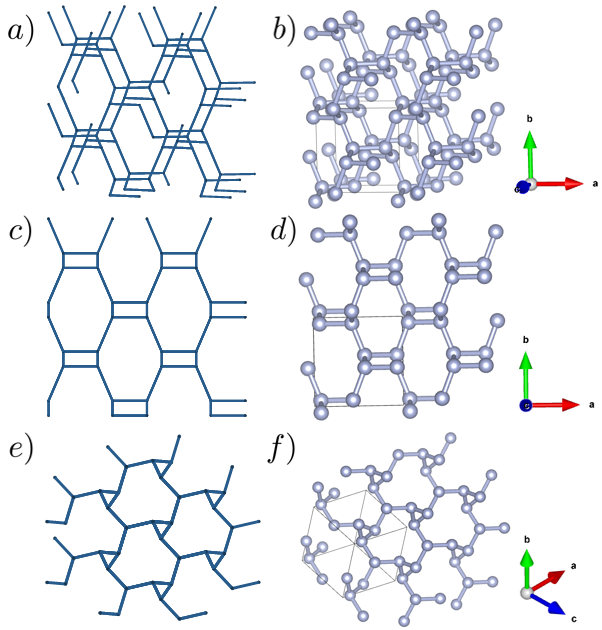}
            \caption{Shown in (a), (c) and (e) are three different views of the structure in Fig.~\ref{fig.Stripes}(c), after including a concomitant distortion. Shown in (b), (d) and (f) are three views of the cubic gauche structure\cite{Laniel2019} from angles similar to the figures on the left. }
            \label{fig.cg_N}
        \end{figure}
        \begin{figure}
            \includegraphics[width=0.8\columnwidth]{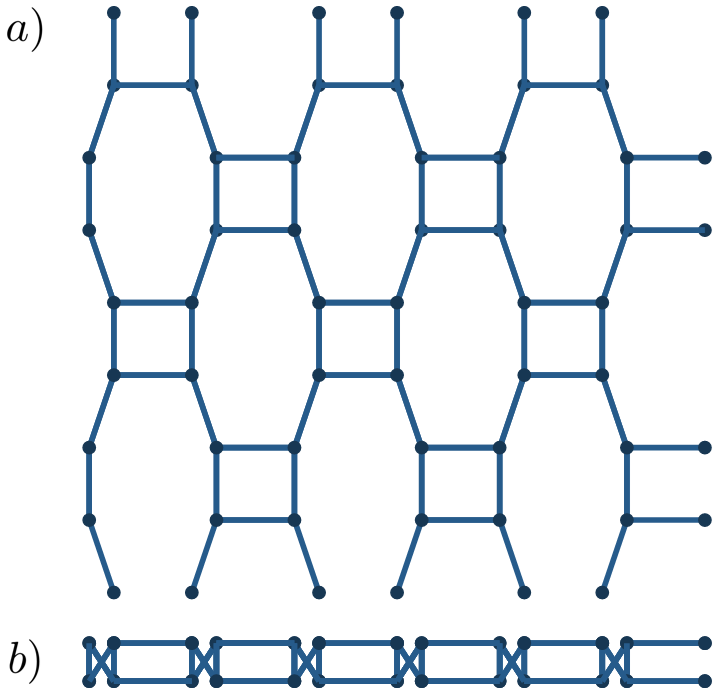}
            \caption{Two views of a single layer of the square-octagon structure in Figs.~\ref{fig.Stripes}(d,f). Sites have been shifted to reflect a concomitant distortion.}
            \label{fig.Stripes_SO}
        \end{figure}
    \end{enumerate}

\section{Black Phosphorus}
\label{sec.BlackPhosphorus}
\begin{figure}
        \includegraphics[width=0.8\columnwidth]{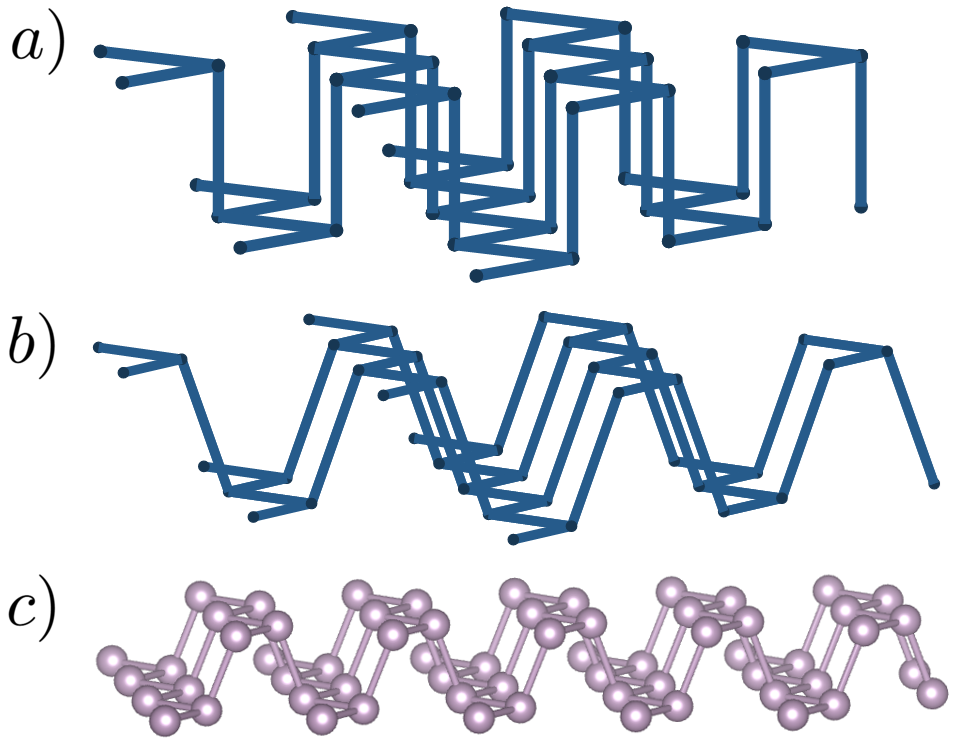}
        \caption{$a)$ Network configuration on a bilayer, when the two boundary Ising chains are in antiferromagnetic configurations. $b)$ The same structure with a structural distortion. Bonds with a dimer have been made shorter, while others are made longer. $c)$ The structure of phosphorene\cite{Hultgren1935}.}
        \label{fig.Phosphorene}
    \end{figure}
    \begin{figure}
    \includegraphics[width=1\columnwidth]{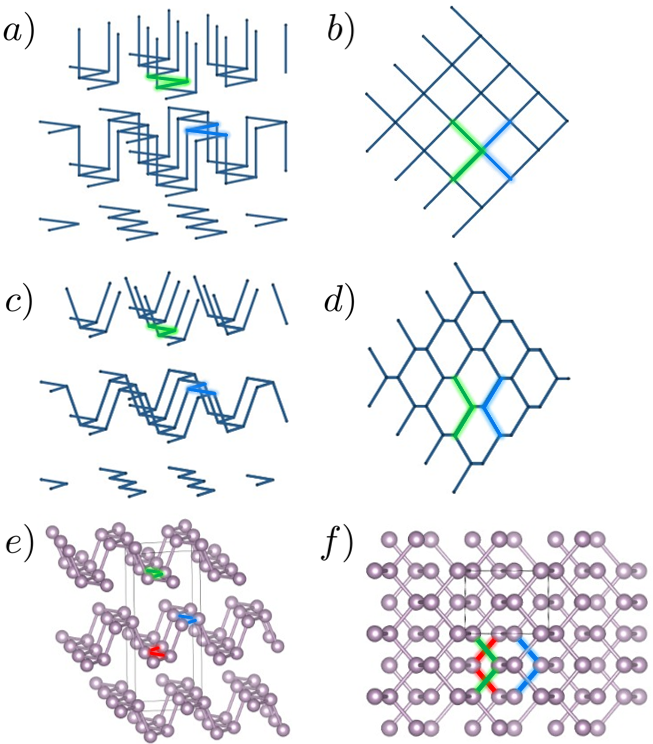}
        \caption{Shown in (a) and (b) are two views of AA-stacked bilayer structures in which each bilayer is in the antiferromagnetic phase. Some dimers in (a) and (b) are highlighted for better visualization. The same structure, with a concomitant distortion, is shown in (c) and (d). 
        The crystal structure of black phosphorus is shown in (e) and (f). By comparing (c,d) with (e,f), we see that bilayers are shifted with respect to one another in black phosphorus.
        }
        \label{fig.BlackPhos}
    \end{figure}

Black phosphorus is a van der Waals material with puckered layers that are held together by weak attractive forces. The layers can be separated, with the single-layer material denoted as phosphorene.

To describe the structure of (black) phosphorene, we modify our starting point from a simple cubic lattice to a square-bilayer. We consider a system with two layers, both parallel to the $xy$ plane. When the rules of our trivalent network model are applied to this geometry, every bond in the $z$ direction must necessarily host a dimer. This can be seen as follows. As argued in Sec.~\ref{sec.configspace}, each site is attached to a tripod with a dimer each along the $x$, $y$ or $z$ axes. With a bilayer geometry, each site is connected to precisely one bond in the $z$ direction. As a result, this bond must necessarily host a dimer. Along $x$ and $y$ directions, as argued before, we will have lines with alternating dimers and blanks. The problem then reduces to four one-dimensional Ising chains defined at the boundaries, two on the $zx$ plane and two on the $yz$ plane (representing lines along $y$ and $x$ respectively). When Ising chains in each plane are in antiferromagnetic phases, we obtain the structure of phosphorene. This is shown in Fig.~\ref{fig.Phosphorene}.

The structure of bulk black phosphorus is denoted as A17. This structure can be obtained by distorting a simple cube, but not as described in previous sections. To obtain this structure, we require a ferromagnetic configuration of Ising variables on one surface and antiferromagnetic configurations on the other two. This goes against the symmetric effective model of Eq.~\eqref{eq.E} whereby all surfaces must have the same Ising ground state. Rather than invoking Eq.~\eqref{eq.E}, we explain the A17 structure as arising from stacking bilayers. That is, we imagine bulk phosphorus as arising from the stacking of pre-formed phosphorene layers.
We show AA-stacked bilayers in the antiferromagnetic phase in Figs.~\ref{fig.BlackPhos}(a,b). The same structure, with a suitable distortion, is shown in Figs.~\ref{fig.BlackPhos}(c,d). Each site is shifted by a small amount towards the base of its tripod (see Fig.~\ref{fig.Tripods}). The resulting structure can be compared with the crystal structure of black phosphorus shown in Figs.~\ref{fig.BlackPhos}(e,f). Visual comparison shows that bilayers must be shifted with respect to one another to recover the structure of black phosphorus. This serves to minimize repulsion between proximate dimers. The shift between layers has also been discussed in previous studies\cite{Dai2014,Cakir2015}.

\section{Impurity textures}
\label{sec.impurities} 

To test the validity of our trivalent network approach, we present the following experimentally testable proposition. When a single impurity is implanted in a pnictogen solid, it will necessarily create one or more long-ranged domain walls. The impurity may be a chalcogen atom, e.g., sulphur in phosphorus or selenium in arsenic. Or, it may be a halogen atom, e.g., fluorine in nitrogen or iodine in antimony. While pnictogen atoms form three valence bonds as described by the trivalency rule, chalcogens and halogens form two and one respectively. In the presence of such an impurity, the trivalency rule will have to be modified locally. We argue that this will necessarily create a long-ranged defect. In the following discussion, we focus on the A7 and A17 structures that are well established in experiments.

\begin{figure}
    \includegraphics[width=1\columnwidth]{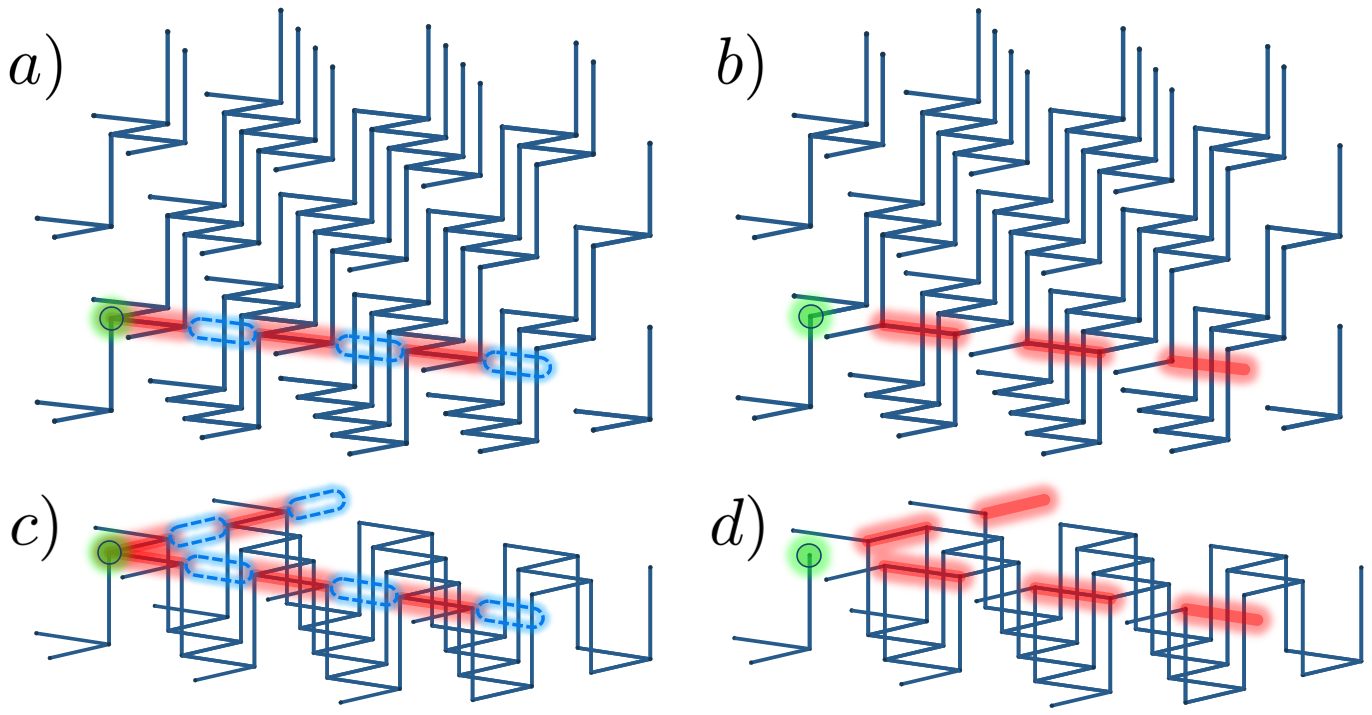}
        \caption{A bulk configuration in the antiferromagnetic phase is shown in (a). A divalent impurity is introduced at the site highlighted in green. To accommodate this impurity, we must shift dimers along a semi-infinite line. These are shown as red dimers that are to move into blank positions shown in blue. The resulting configuration is shown in (b). 
         A bilayer in the antiferromagnetic phase is shown in (c).
        A monovalent impurity is introduced at the green site. This necessitates shifting dimers on two semi-infinite lines. The resulting configuration is shown in (d).
        }
        \label{fig.Impurities}
    \end{figure}

We first consider the bulk A7 structure, obtained as the antiferromagnetic phase of the boundary Ising models. As seen from Fig.~\ref{fig.J1AntiF}, this structure naturally forms separable layers. When a chalcogen impurity is introduced (e.g., tellurium atoms in antimony), the impurity can only have two bonds and not three. In order to satisfy this requirement, we must shift dimers along a semi-infinite line that emanates from the impurity site as shown in Fig.~\ref{fig.Impurities}(a,b). Crucially, this necessarily creates bonds across the initially-separable layers. As a consequence, a single impurity can make the structure more rigid. A small number of impurities can stiffen the structure, making it difficult to separate into layers.

We next consider the structure of phosphorene, a bilayer structure with antiferromagnetic Ising configurations. We consider a single halogen impurity as shown in Fig.~\ref{fig.Impurities}(c). The impurity site can only have a single dimer attached. As a result, we must shift bonds along two lines that emanate from the impurity. This separates the material into multiple grains, with the grain boundaries meeting at the impurity site. We argue that this is a generic feature of all structures obtained from our trivalent network model. A small number of impurities will create domains and domain walls that can be detected, e.g., by using scanning tunneling microscopy.

\section{Discussion}
\label{sec.discussion}
    We present an effective model to describe the structures of group-15 materials. We describe these materials as covalently-bonded solids, akin to diamond. Covalent bonds arise from stronger hopping amplitudes on bonds that are shortened by distortion. With three electrons on each atom, inter-orbital repulsion is inevitable. However, intra-orbital repulsion, which is stronger, can be avoided. This is achieved by a bending constraint, which leads to a bulk-boundary mapping and a simple boundary theory. Future studies could revisit our results from a microscopic point of view, including effects of Hund's coupling and spin-orbit coupling.
    The covalent-bonding picture is fully consistent with the insulating nature of materials such as cubic-gauche nitrogen\cite{Chen2008} and black phosphorus. Solid arsenic, antimony and bismuth are known to form semi-metals with small Fermi surfaces and indirect band gaps\cite{Gonze1990,Behera2023}. Studies can examine whether semi-metallic nature can arise from the excitations of a trivalent covalently-bonded network.

    The trivalent network model provides a framework to explain the A7, A17 and cubic gauche structures seen in pnictogen solids. It can also be adapted to explain the layered structures of antimonene, black phospherene and the proposed square-octagon layer. The proposed cubic molecule, N$_8$, may also derive from a phase of the trivalent network model. 
    There are a few other structures of group-15 elemental solids that are not captured by the trivalent network model. For example, nitrogen forms two-dimensional polymeric structures that cannot be described as deformations of a cubic structure\cite{Ma2009}. Network models may be able to describe these structures by starting from idealized structures other than the simple cubic lattice.

    The trivalent network model has a configuration space that grows exponentially with surface area, rather than volume, as shown in Eq.~\eqref{eq.N_c}. This feature is shared by models proposed in the context of fractonic particles, where the ground state degeneracy grows in a sub-extensive fashion\cite{Slagle2017,Nandkishore2019,Gromov2024}. In fact, a valence-cube-solid model, similar to the  breathing cube structure of Fig.~\ref{fig.BreathingCubes}, has been proposed as a host for fractonic excitations\cite{You2020}. Hole doping can produce divalent defects (with two dimers rather than three). Such defects, in isolation, are immobile as they cannot move without violating constraints elsewhere. However, if four such defects are placed on a square face, this arrangement can move perpendicular to the plane of the square. Similar arguments  hold in other phases of the trivalent network model. For example, Fig.~\ref{fig.Impurities}(b) depicts a point-like defect in the A7 structure. This defect can easily move in one direction (by shifting one or more red dimers), but not along others. 
    Our results suggest that fractonic excitations can be realized in group-15 elemental solids. This could make fractons accessible to solid-state probes, e.g,. to transport measurements such as conductivity anisotropy.

    Within our effective boundary theory, structures such as A7, A17 and cubic gauche are found in distinct parameter regimes. An exciting future direction is to determine the parameters of the boundary theory ($J_1$ and $J_2$) from microscopic considerations. They will receive contributions from Coulomb interactions as well as the mechanical energy cost of inducing distortions.  
    Microscopic estimates could explain why a given material takes on certain structures, but not others. It could also clarify the role of pressure and temperature which serve as macroscopic tuning parameters\cite{Hogan2021,Laniel2020,Ji2020,Li2018}. A microscopic approach may also suggest relevant corrections to our minimal model. For example, within our model, the cubic gauche structure is degenerate with the square-octagon structure of Figs.~\ref{fig.Stripes}(d,f). Nitrogen is known to exhibit the former structure, but not the latter. This indicates that further corrections, possibly multi-dimer interactions, disfavour four-dimer rings.

    \acknowledgments
    We thank Han Yan for helpful discussions. This work was supported by National Sciences and Engineering Research Council of Canada (NSERC) through Discovery Grant 2022-05240. RG thanks the Office of Global Engagement, IIT Madras  for hospitality and support.

\appendix

    \setcounter{figure}{0}
    \setcounter{equation}{0}
    \renewcommand{\theequation}{A\arabic{equation}}
    \renewcommand{\thefigure}{A\arabic{figure}}
    
    \section{Unique Structures in the Stripe Phase}
    \label{app.lines}

    \begin{figure}
        \includegraphics[width=1\columnwidth]{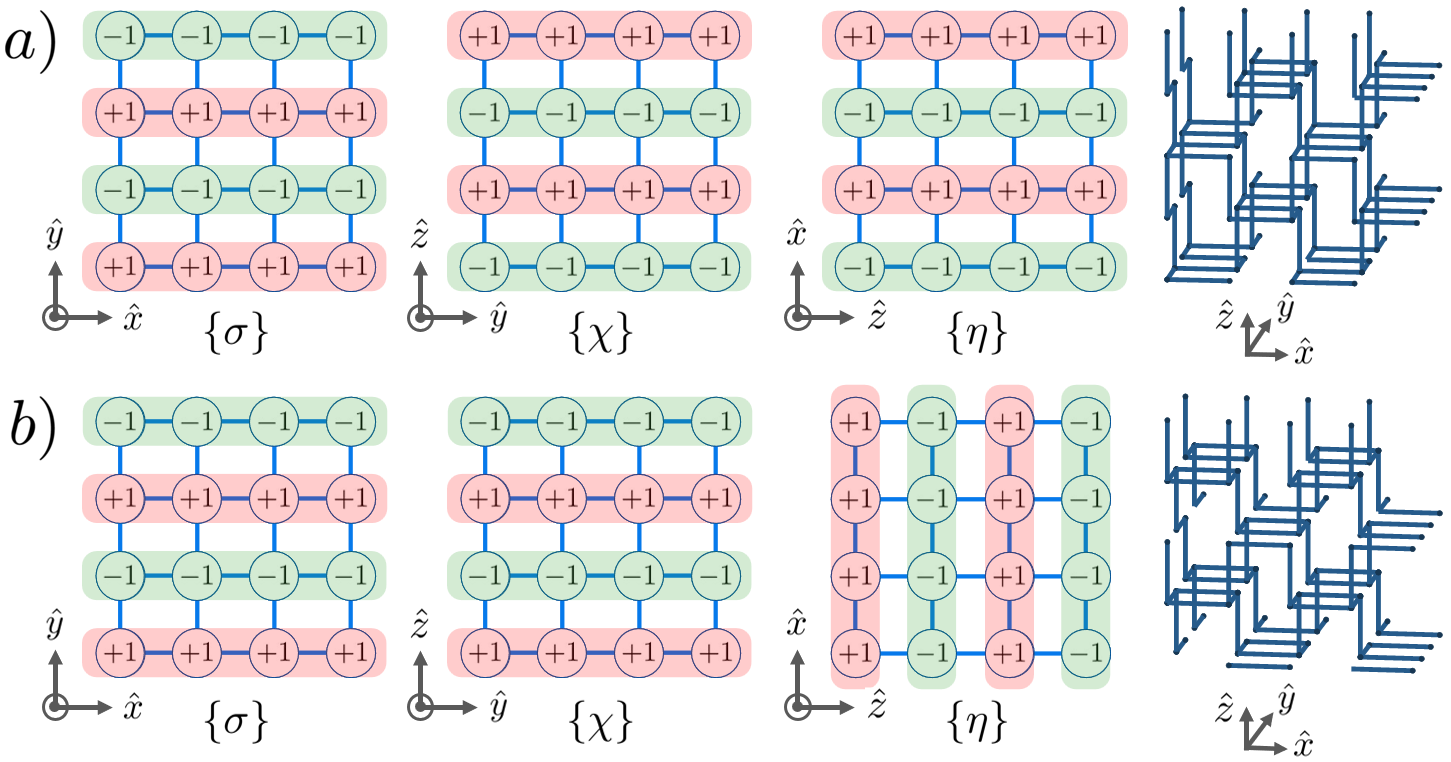}
        \caption{$a)$ and $b)$ show the patterns of stripes formed by Ising variables on the three bounding surfaces of a $4\times4\times4$ cubic lattice. The resulting bulk crystal structure is shown on the right. The top row depicts the cubic gauche structure of Figs.~\ref{fig.Stripes}(c,e). The bottom row shows the square-octagon layered structure of Figs.~\ref{fig.Stripes}(d,f). }
        \label{fig.Stripes_app}
    \end{figure}
    As discussed in Sec.~\ref{sec.phasediagram}, the stripe phase of the boundary theory yields two distinct bulk crystal structures.    Fig.~\ref{fig.Stripes_app} shows boundary configurations that yield these structures on a $4\times4\times 4$ cubic lattice. The patterns can be easily extended to larger system sizes. Symmetry-related copies of these structures can also be obtained from the boundary theory.

    \bibliography{main}

\end{document}